\documentclass{jfm}
\usepackage{graphicx}
\usepackage{epstopdf, epsfig}
\usepackage{textcomp}

\shorttitle{Guidelines for authors}
\shortauthor{A. N. Other, H.-C. Smith and J. Q. Public}

\title{Confinements regulate capillary instabilities of fluid threads}

\author{Xiaodong Chen\aff{1,2},
       Chundong Xue\aff{1,3},
       \and Gongqing Hu\aff{1,3,}\corresp{\email{guoqing.hu@imech.ac.cn}}}

\affiliation{\aff{1}State Key Laboratory of Nonlinear Mechanics, Institute of Mechanics, 	Chinese Academy of Sciences, Beijing 100190, China	
\aff{2}School of Aerospace Engineering, Beijing Institute of Technology, Beijing 100081, China
\aff{3}School of Engineering Science, University of Chinese Academy of Sciences, Beijing 100049, China
}
\begin{document}

\maketitle

\begin{abstract}
We study the breakup of confined fluid threads at low flow rates to understand instability mechanisms. To determine the critical conditions between the earlier quasi-stable necking stage and the later unstable collapse stage, simulations and experiments are designed to operate at an extremely low flow rate. Critical mean radii at neck centres are identified by the stop-flow method for elementary microfluidic configurations. Analytical investigations reveal two distinct origins of capillary instabilities. One is the gradient of capillary pressure induced by the confinements of geometry and external flow, whereas the other is the competition between local capillary pressure and internal pressure determined by the confinements.
\end{abstract}


\section{Introduction}
Fluid thread breakup is a widespread phenomenon in nature, industry, and daily life \citep{Eggers2007,Bhat2010,VanHoeve2010,Tagawa2012}. As a common example, columnar water jets from a showerhead are unstable downstream of the nozzles and break into droplets in air. This breakup scenario is called Rayleigh-Plateau (R-P) instability \citep{Plateau1873,Rayleigh1878}, which is also called capillary instability at low flow rates where the driving force is interfacial tension \citep{Eggers2007,Eggers2008}. In contrast, a fluid thread covered by a thin film of another fluid within a tube can be stable \citep{Duclaux2006}. For instance, a long air bubble in an intravenous line maintains its shape while moving slowly with the liquid. This kind of tightly confined fluid thread is often encountered in various processes, including pore-scale breakup in enhanced oil recovery \citep{Olbricht1996} and droplet or bubble generation in microfluidic devices \citep{Squires2005,Anna2016}, where confined fluid threads are squeezed into fluid necks at channel junctions and eventually break into microdroplets. The interface is quasi-stable in the earlier necking stage and becomes unstable in the later collapse stage \citep{Garstecki2005}. The stage transition is thought to be an R-P instability \citep{Link2004,Garstecki2005,DeMenech2008} or initiated not by an R-P instability but rather by a reversed flow from the thread tip to the neck \citep{VanSteijn2009}. The mechanisms of the onset of interfacial instabilities are still puzzling.

Here, through newly designed numerical simulations and experiments at extremely low capillary numbers, we identify transitions between the quasi-stable and unstable stages for confined fluid threads. Two breakup mechanisms are uncovered from theoretical analyses in surprisingly simple terms.

\section{Problem statement}
\begin{figure}
  \centerline{\includegraphics[width=12cm]{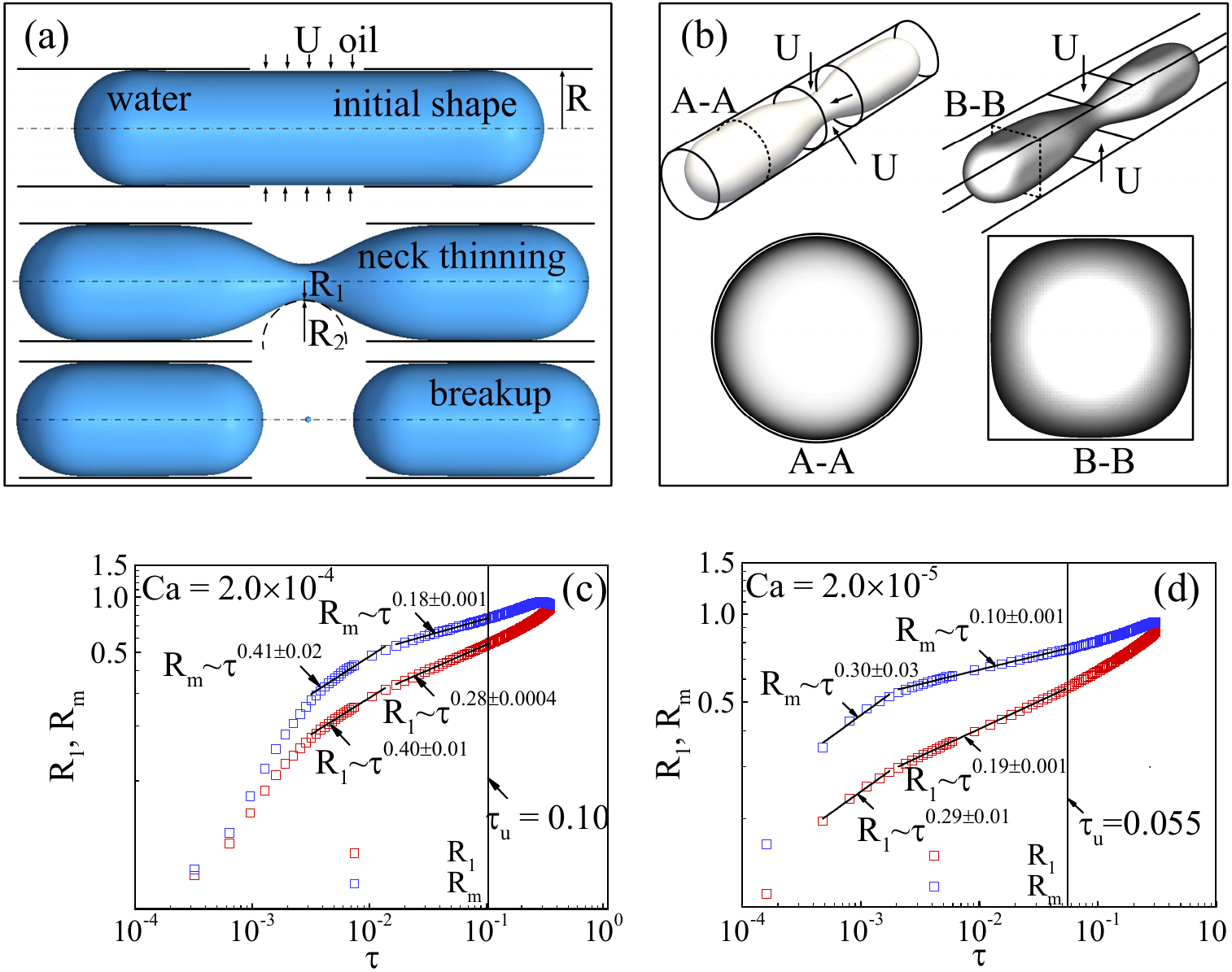}}
  \caption{Axisymmetric breakup of a confined fluid thread. (a) Sketch of the breakup process. (b) Comparison of droplet shapes in channels with circular and rectangular cross-sections during breakup. (c) and (d) Time evolutions of $R_1$ and $R_m$ for $Ca (\mu_wU/\sigma)$ of $2.0\times10^{-4}$ and $2.0\times10^{-5}$, respectively.}
\label{fig:1}
\end{figure}

Various microfluidic configurations are considered in this paper. Inspired by the axisymmetric configurations used in studying the breakup mechanism of unbounded fluid threads \citep{Gordillo2005,Eggers2007}, we first investigate the breakup of an axisymmetric water thread that is tightly confined inside a circular microchannel, as shown in figure \ref{fig:1}(a). A fluid neck forms under the squeezing of the oil side flow and then collapses. As illustrated in figure \ref{fig:1}(b), a circular channel avoids the empty corners in a rectangular channel where oil inevitably flows through the corners to cause leakage and thus makes it difficult to deform the interface under low flow rates \citep{Dollet2008,Shui2008}. Using circular channels, we can overcome the leakage problem and study the breakup at extremely low flow rates. Non-axisymmetric configurations, including a T-junction configuration with circular microchannels and configurations with square microchannels, are also studied and are described when mentioned.

\section{Numerical and experimental methods}
Numerical simulations are conducted using the Gerris flow solver (http://gfs.sf.net). Combining an adaptive quad/octree spatial discretization, geometrical volume-of-fluid interface representation, balanced-force continuum-surface-force surface tension formulation, and height-function curvature estimation, the method is shown to recover exact equilibrium (to machine accuracy) between surface tension and pressure gradient in the case of a stationary droplet, irrespective of viscosity and spatial resolution \citep{Popinet2009}. The accuracy of the code has been validated in our previous studies for various interfacial flow problems \citep{Chen2013,Chen2014}. A thickness-based refinement criterion developed in our previous work \citep{ChenYang2014} is added to resolve the thin film between the interface and the wall.

Experiments are performed in circular microchannels that are fabricated by a rapid prototyping method \citep{Ghorbanian2010}. Iron wires are cut and arranged on a cured polydimethylsiloxane (PDMS) flat piece. Wires are connected by dropping a small amount of melted paraffin in gaps between the wires. The connected wires are then covered with the uncured PDMS. After the PDMS is cured, the wires are pulled out to form the desired configurations. More experimental details can be found in the Supplementary Material.

\section{Results}
\subsection{Neck evolutions during the axisymmetric breakup}
Numerical simulations are first carried out for the axisymmetric configuration in figure \ref{fig:1}(a). A water droplet (viscosity $\mu _w$ = 1 mPa·s, density $\rho _w$ = 998 kg/m$^3$) is initially tightly confined in a circular microchannel with radius $R$ = 25 \textmu m. The initial shape of the droplet is a cylinder (radius $R$ = 25 \textmu m) with two hemispherical tips. The incoming oil (viscosity $\mu _o$ = 8 mPa·s, density $\rho _o$ = 770 kg/m$^3$) from the lateral opening (width of 50 \textmu m) deforms the interface (interfacial tension $\sigma$ = 5 mN/m). A 2-D axisymmetric model is established to consider only a quarter of the cross-section of the configuration. Neck evolutions during the axisymmetric breakup of the confined fluid thread are characterized by two principal radii at the neck centre, $R_1$ and $R_2$, in the radial and axial directions, respectively, and the mean radius $R_m = R_1R_2/(R_1 + R_2)$ according to the Young-Laplace equation.

Figure \ref{fig:1}(c) and (d) shows the evolutions of $R_1$ and $R_m$ for capillary number $Ca = \mu_w U/\sigma = 2.0\times10^{-4}$ and $2.0\times10^{-5}$, respectively, where $\tau = t_c - t$ represents the time remaining until pinch-off, with $t$ the time and $t_c$ the time at pinch-off. The radii are normalized by $R$, and time is scaled by $R/U$. In figure \ref{fig:1}(c), we identify two stages where $R_1$ and $R_m$ exhibit power law behaviours with $\tau^{\alpha}$, where the scaling constant $\alpha$ is a signature of the physical mechanisms causing the collapse \citep{VanHoeve2011}. The values of $\alpha$ in the first and second (based on $t$) collapse stages are 0.28$\pm$0.0004 and 0.40$\pm$0.01, respectively, which agrees well with the experimental measurements and theoretical predictions of 1/3 and 2/5 for microbubble formation in a microfluidic flow-focusing device \citep{VanHoeve2011}. The values of $\alpha$ for $R_m$ of the two stages are 0.18$\pm$0.001 and 0.41$\pm$0.02, respectively, suggesting that the effect of $R_2$ on $R_m$ is not negligible in the first collapse stage and the earlier necking stage. It is therefore more reasonable to use $R_m$ to characterize the earlier necking stage, rather than $R_1$ that has been widely used to characterize the collapse stages \citep{VanHoeve2011,CastrejonPita2015}. Notably, complex scaling behaviours emerge after the second collapse stage. Several transient stages may exist until pinch-off owing to the formation of satellite droplets at the neck centre \citep{CastrejonPita2015}. Similar behaviours of $R_1$ and $R_m$ are observed at an extremely low $Ca$ of $2.0\times10^{-5}$, as shown in figure \ref{fig:1}(d). The values of $\alpha$ in the two stages are smaller than those of $Ca = 2.0\times10^{-4}$. In the earlier necking stage before the first collapse stage for both $Ca$ numbers, variations in the radii do not conform to any power law with a constant $\alpha$. We thus expect a critical condition representing the transition between the earlier necking stage and the later collapse stages.

\subsection{Critical mean radius for the axisymmetric breakup}
\begin{figure}
  \centerline{\includegraphics[width=12cm]{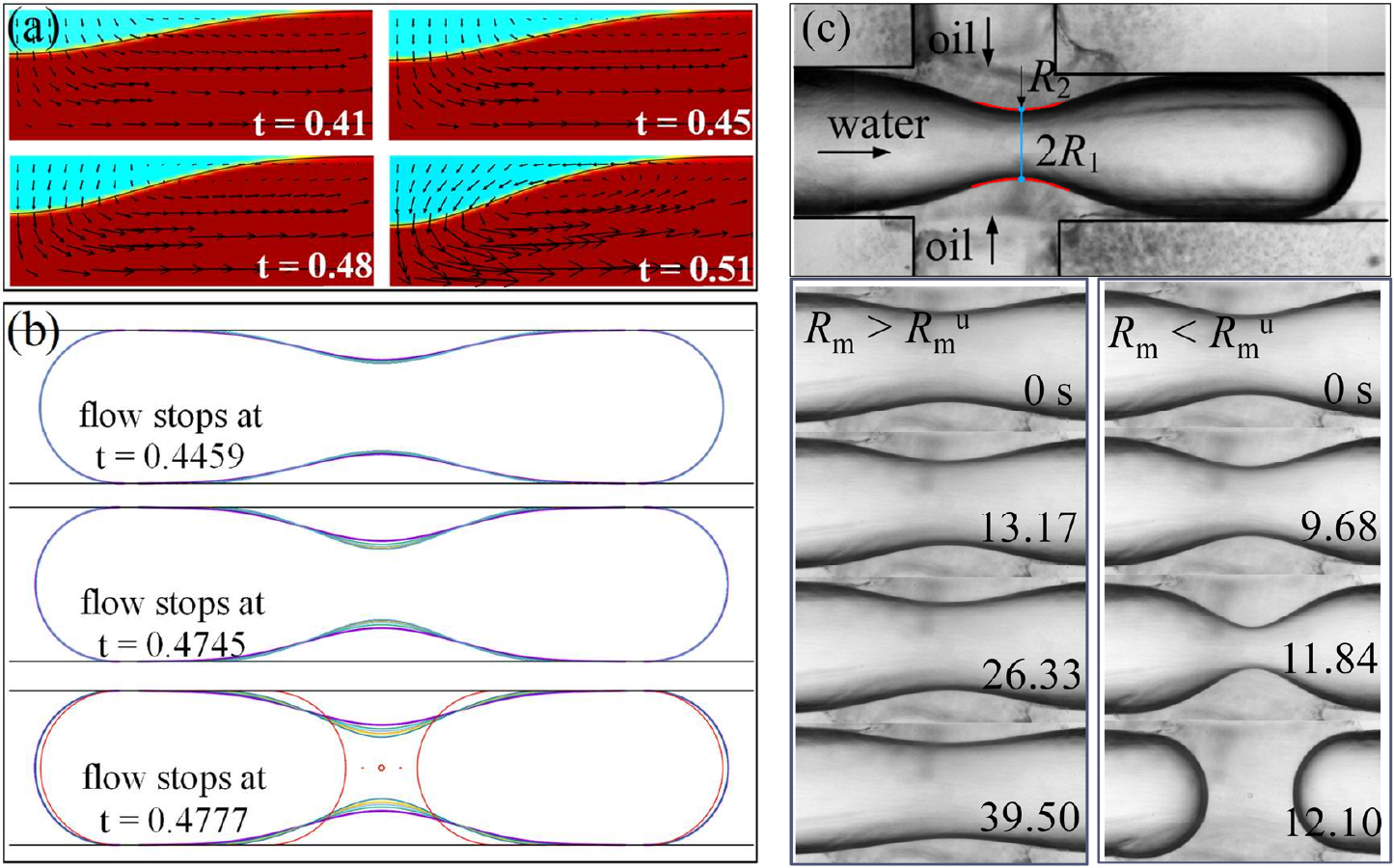}}
  \caption{Determination of the critical mean radii $R_m^u$ at the neck centre for axisymmetric configurations. (a) Transition of the flow pattern during breakup for $Ca = 2.0\times10^{-5}$. (b) Evolution of the interface during stop-flow simulations. The initial shapes are adapted from the simulation for $Ca = 2.0\times10^{-5}$. (c) Stop-flow experiments to determine $R_m^u$. Videos for $R_m > R_m^u$ and $R_m < R_m^u$ are available in the Supplementary Material.}
\label{fig:2}
\end{figure}

From the evolution of the flow field, we find a transition of the flow pattern near the neck. As shown in figure \ref{fig:2}(a), oil flows expansively to squeeze the interface at $t$ = 0.41 and 0.45, while oil flows towards the neck centre at $t$ = 0.48 and 0.51. Here, the critical time $t_u$ when the transition occurs can be determined by the stop-flow method \citep{Stone1986,Blanchette2006,Hoang2013}. The flow is stopped at different moments by setting the flow velocity to zero and then the simulations resume. We find that, at $t_u$, the critical mean radius $R_m^u$ is $(0.76\pm0.002)$ for $Ca = 2.0\times10^{-5}$. As shown in figure \ref{fig:2}(b), if the flow stops when $R_m > R_m^u$, the interface remains stationary after small relaxations at $t$ = 0.4459 and 0.4745. If flow stops when $R_m < R_m^u$, the neck thins spontaneously until it collapses (as at $t = 0.4777$). The mean radius of the stationary interface corresponding to $R_m^u$ is approximately $0.72$. The critical remaining time $\tau_u = t_c - t_u$ is obtained for each $Ca$ number in figure \ref{fig:1}(c) and (d). $\tau_u$ preferably represents the duration of the late collapse stage, indicating that the transition of the flow pattern is equivalent to the transition between the earlier necking stage and the later collapse stage. We also obtain $R_m^u$ for two different fluid systems with different viscosity ratios and interfacial tensions for $Ca$ numbers down to $2.5\times10^{-6}$ (see Supplementary Material). There are small distinctions in $R_m^u$ among the different fluid systems. The mean neck radii of the stationary interfaces corresponding to $R_m^u$, however, are in good agreement. The physical properties of the fluids thus have negligible effects on the critical neck shape at low $Ca$ numbers.

The existence of the transition is further confirmed by stop-flow experiments, shown in figure \ref{fig:2}(c), using a flow-focusing microfluidic device with circular channels, where $\mu_w$ = 1 mPa·s, $\rho_w$ =998 kg/m$^3$, $\mu_o$ = 20 mPa·s, $\rho_o$ = 950 kg/m$^3$, $\sigma$ = 35 mN/m, and $R$ = 400 \textmu m. The water flow is switched off after the thread tip passes the channel junction. The oil flow is stopped and resumed intermittently to determine the critical shape of the neck. $R_m^u$ is found to be $(0.79\pm0.04)$, in agreement with the value of $(0.76\pm0.002)$ obtained from simulations. When the flow stops at $R_m > R_m^u$, the interface relaxes inwards, as in the simulations, and then slowly moves outwards. This outward motion may be attributed to the leakage of oil from the neck region through the gap between the rough channel wall and the interface. When the flow stops at $R_m < R_m^u$, the neck collapses irreversibly, as in the simulations.

The above stop-flow studies provide direct evidence that the earlier necking stage is quasi-stable \citep{Garstecki2005}. Our simulations suggest that the internal pressure of the droplet is nearly constant during the quasi-stable necking stage. Using the pressure outside of the droplet tips as the reference pressure, the internal pressure of the fluid thread can be estimated by the capillary pressure of the tips, $\sigma/R_m$, which equals $2\sigma/R$ since $R_m = R/2$ ($R_1 = R_2 = R$ at the tips). Additionally, the oil pressure outside the neck is almost evenly distributed during the quasi-stable stage. The capillary pressure is thus nearly constant along the neck interface, indicating that $R_m$ can be considered constant along the neck region. In the axisymmetric configuration, the neck interface is essentially an axisymmetric surface of constant mean curvature (CMC) or a Delaunay surface \citep{Delaunay1841}. $R_m$ varies from $R$ to approximately $0.76R$ in the quasi-stable stage. The corresponding capillary pressure, varying from $\sigma/R$ to $1.32\sigma/R$, is consistently smaller than the internal pressure, $2\sigma/R$. The internal pressure pushes the interface outwards to stabilize it against inward collapse. The stabilization of the confined fluid thread thus originates from the domination of the internal pressure caused by the thread tips under confinement.

\subsection{Destabilizing mechanism of the axisymmetric fluid thread}
\begin{figure}
  \centerline{\includegraphics[width=12cm]{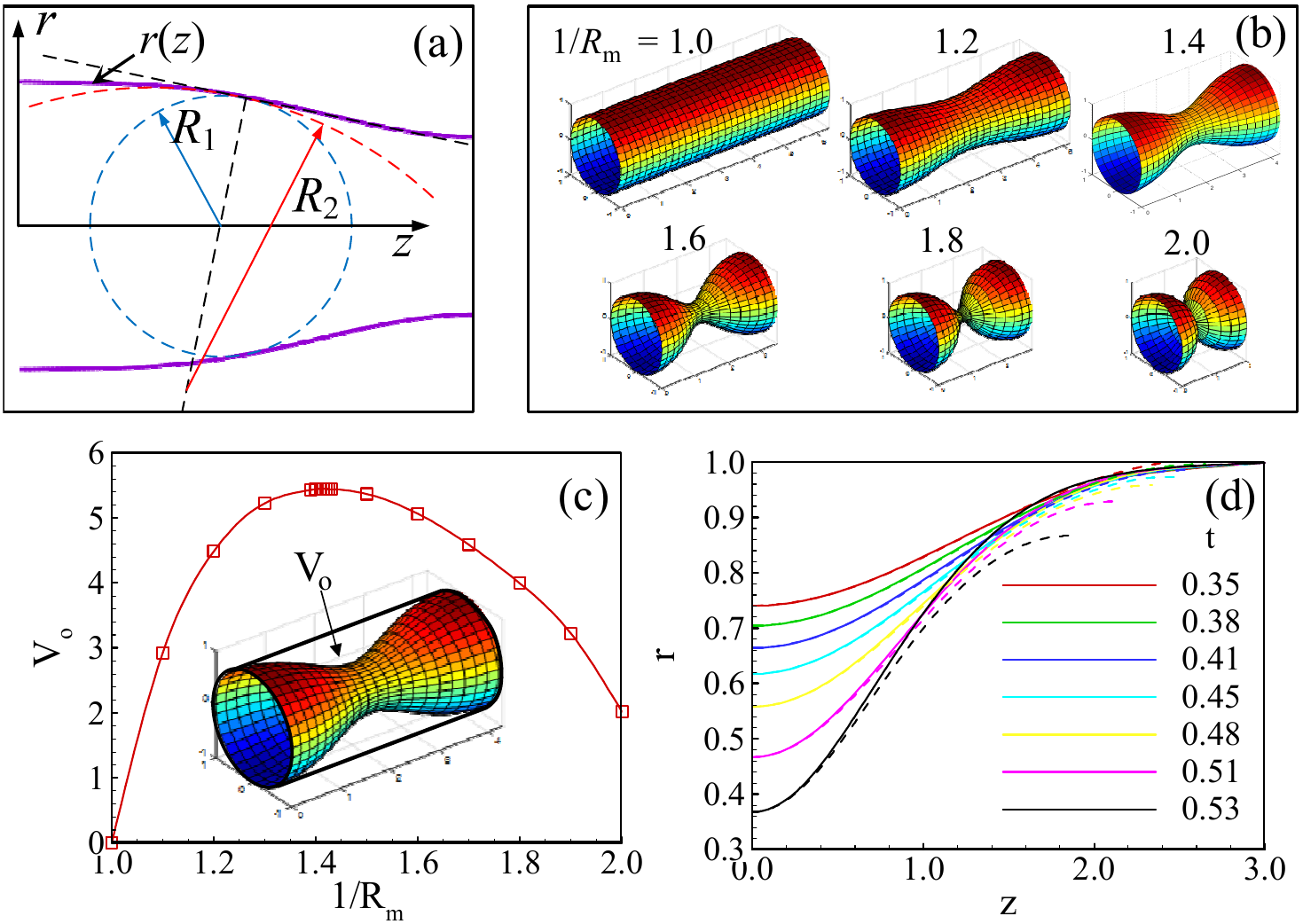}}
  \caption{Theoretical demonstration and validation of the existence of the critical $R_m$. (a) Sketch of the neck profile consisting of a curve of CMC. (b) Axisymmetric CMC surface with increasing mean curvature. (c) Evolution of the oil volume between the interface and the cylindrical channel with the mean curvature for the CMC surface. (d) Comparison of the neck profiles between numerical simulations (solid lines) and CMC curves (dashed lines) using the neck radii from the simulations as boundary conditions.}
\label{fig:3}
\end{figure}

The CMC feature of the neck interface inspires us to propose a theoretical analysis to reveal an unrecognized destabilizing scenario of capillary instability. The two principal radii are derived from the geometrical relationships

\begin{equation}
{R_1} = r(z){[1 + \dot r{(z)^2}]^{1/2}},
\end{equation}

\begin{equation}
{R_2} =  - \frac{{{{[1 + \dot r{{(z)}^2}]}^{3/2}}}}{{\ddot r(z)}}
\end{equation}

\noindent where $r(z)$ is the profile of the neck in figure \ref{fig:3}(a). The shapes of the neck for $R \le 1/R_m \le 2/R$ are obtained by solving $1/R_m = 1/R_1 + 1/R_2$ with boundary conditions of $r(0) = R$ and $\dot{r}(0) = 0$ by a MATLAB code.

Figure \ref{fig:3}(b) shows that the length of the neck region decreases near-linearly with $1/R_m$. The initial cylindrical shape has the maximum length, 2{\rm$\pi$}$R$, according to Delaunay’s method \citep{Delaunay1841}. Figure 3(c) shows that the oil volume between the neck interface and the columnar channel wall first increases and then decreases with increasing $1/R_m$. During the whole process of necking, oil inevitably flows into the channel, resulting in the oil volume increasing with $1/R_m$ during the breakup. When the volume is larger than the maximum allowed for the CMC surfaces, $R_m$ can no longer be constant along the neck interface. A capillary pressure gradient thus emerges along the interface to promote an irreversible collapse. Figure \ref{fig:3}(c) shows that the maximum oil volume for the CMC surfaces occurs at $R_m = (0.71\pm0.001)R$, agreeing with the stationary neck shape obtained numerically for $Ca = 2.5\times10^{-6}$.

We further support this finding directly by comparing the simulated neck profiles for $Ca = 2.0\times10^{-5}$ with the corresponding CMC curves. Figure \ref{fig:3}(d) shows that good agreement is achieved before the critical moment tu. The destabilization of axisymmetrically confined fluid threads is clearly the result of the gradient of capillary pressure induced by the confinements of both geometry and flow. This instability scenario can be termed confinement-induced capillary instability.

\subsection{Destabilizing mechanism of a non-axisymmetric fluid thread}
\begin{figure}
  \centerline{\includegraphics[width=12cm]{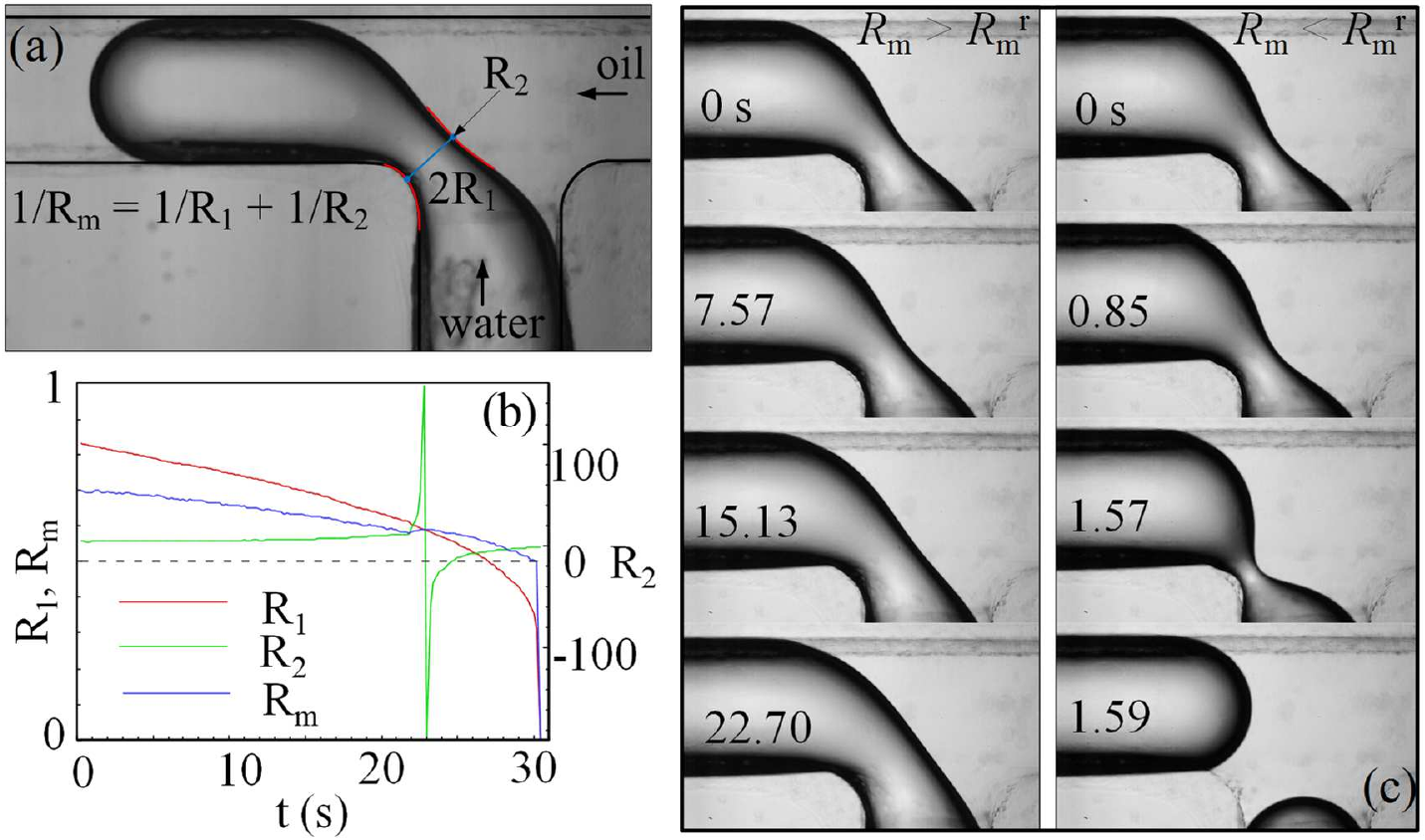}}
  \caption{Rapid collapse of the neck region and determination of the critical mean radius at the neck centre $R_m^r$. (a) Experimental configuration and measurements of two principal radii of curvature for fluid thread breakup in a T-junction of a cross-flowing configuration. The physical parameters are the same as those in figure 2(c). (b) Time evolution of the radii at the neck centre for $Ca = 2.0\times10^{-5}$. (c) Interfacial evolution before and after the critical moment from experiments using the stop-flow method. Videos for $R_m > R_m^r$ and $R_m < R_m^r$ are available in the Supplementary Material.}
\label{fig:4}
\end{figure}

We also experimentally study the breakup of a non-axisymmetric fluid thread with circular channels in a cross-flowing configuration, as shown in figure \ref{fig:4}(a). Figure \ref{fig:4}(b) shows the evolution of three radii for $Ca = 2.0\times10^{-5}$. The collapse occurs more rapidly than that in the axisymmetric configuration, suggesting the existence of a different destabilization mechanism. The stop-flow experiment in figure \ref{fig:4}(c) shows that the critical mean radius $R_m^r$ is $(0.51\pm0.01)R$. The capillary pressure at the neck centre is approximately $2\sigma/R$, which is equal to the internal pressure of the fluid thread. The fluid thread thus becomes unstable when the capillary pressure at the neck centre balances the internal pressure.

Interestingly, this mechanism is similar to the classic “snap-off” mechanism \citep{Roof1970} that is identified in porous media approximately half a century ago. Snap-off is associated with the diameter variation in the porous channel \citep{Beresnev2009}, and a similar mechanism is also found in a step-emulsification microfluidic device with rectangular channels \citep{Dangla2013}.

\subsection{Critical conditions for the fluid thread in square microchannels}
\begin{figure}
  \centerline{\includegraphics[width=11cm]{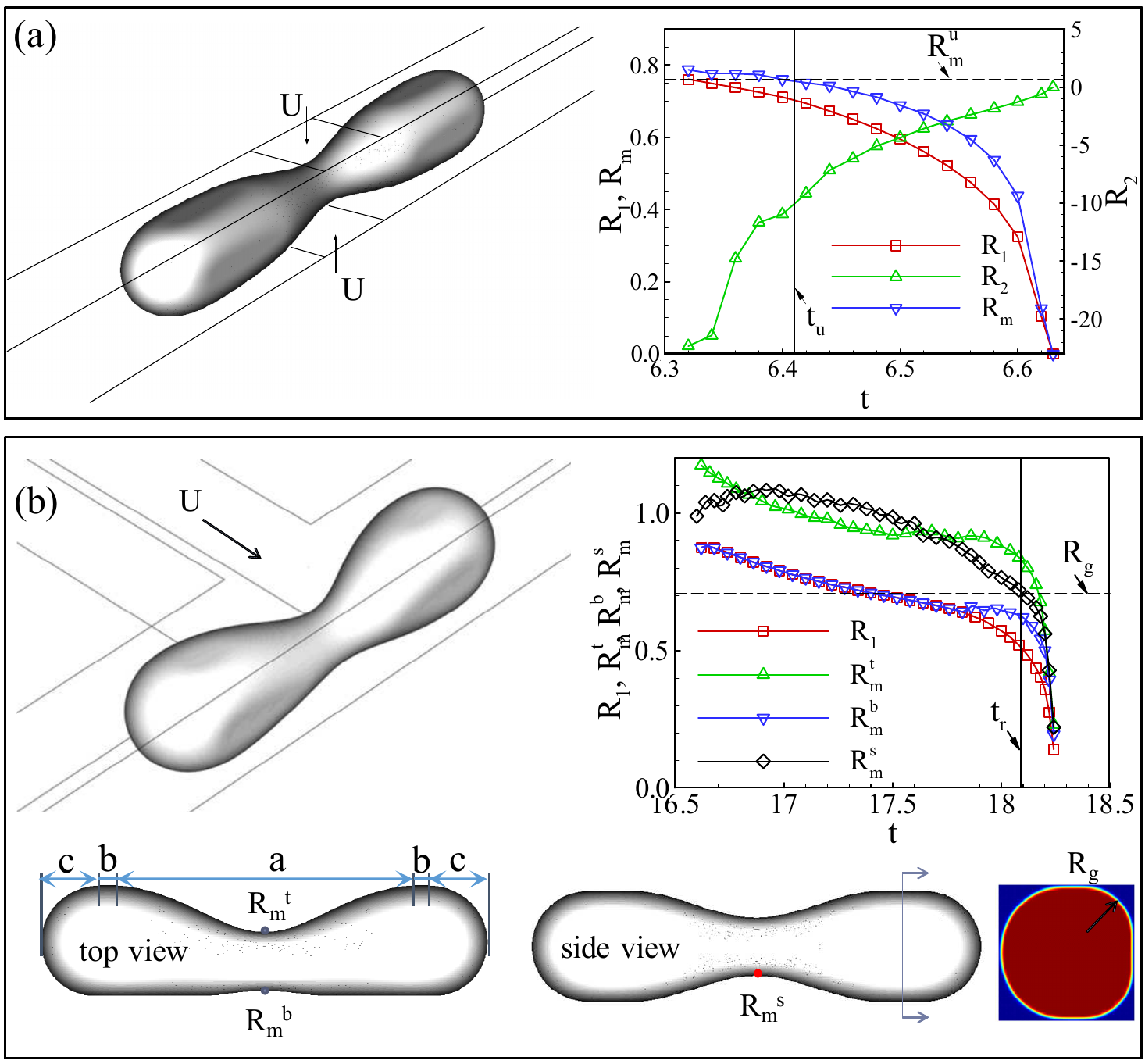}}
  \caption{Determination of the critical mean radii at the neck centre for different configurations with square cross-sections. The channel height $H$ is 25 \textmu m. The physical parameters are the same as those in figure \ref{fig:1}. (a) Droplet breakup in a configuration with symmetrical oil inlets. The left image shows the sketch of the simulation setup. The right plot shows the time evolution of the radii for $Ca = 6.2\times10^{-4}$. The radii are scaled with $H/2$, and time is scaled with $H/2U$. The critical shape is determined by stop-flow simulations. (b) Droplet breakup in a cross-flowing configuration. The top-left image shows the sketch of the simulation setup. The top-right plot shows the time evolution of the radii for $Ca = 6.2\times10^{-4}$. The bottom images show the critical droplet shape determined by stop-flow methods.}
\label{fig:5}
\end{figure}

For completeness, critical conditions for fluid thread breakup in rectangular microchannels are studied numerically. For droplet breakup in a flow-focusing configuration made of square microchannels with height $H$ shown in figure \ref{fig:5}(a), stop-flow simulations demonstrate that $R_m^u = (0.76\pm0.01)(H/2)$, agreeing with that in the axisymmetric configuration. The confinement-induced capillary instability is thus also applicable in square microchannels. For droplet breakup in a T-junction shown in figure \ref{fig:5}(b), we obtain the evolutions of the mean radii at three key positions of the neck centre. The top and bottom radii, $R_m^t$ and $R_m^b$, are influenced by the asymmetric flow. The capillary instability occurs when the mean radius from the side view, $R_m^s$, is approximately equal to the mean radius at the corners, $R_g$, which determines the internal pressure of the droplet \citep{Wong1995}. Here, $R_m^s$ and $R_g$ are $(0.70\pm0.02)(H/2)$ and $0.71(H/2)$, respectively. The origin of the instability is therefore the same as that of the non-axisymmetric case shown in figure \ref{fig:4}, i.e., the fluid thread becomes unstable when the capillary pressure at the neck centre balances the internal pressure.

We also simulate the breakup of a fluid thread in a cross-flowing configuration with rectangular microchannels (see Supplementary Material). We do not observe corner flow from the thread tip to the neck to trigger collapse \citep{VanSteijn2009}. We find only backward flow near the neck region, and the accelerating collapse of the neck region drives the oil phase towards the neck centre. Moreover, breakup occurs even when there is no corner at all, as shown for circular channels in figure \ref{fig:4}.

\section{Discussion}
The above analyses show that the critical conditions for the onset of capillary instability for small $Ca$ numbers are geometrical for confined fluid threads. This finding is consistent with previous experiential studies \citep{Link2004,Jullien2009} on droplet breakup in a T-junction. The boundary between no breakup and breakup for small $Ca$ numbers was found to be independent of $Ca$ number and could be characterized by the critical length of the initial plug-shaped droplet \citep{Jullien2009}. A droplet shorter than the critical length obstructs the junction first, then opens a tunnel on one side of the junction by a random disturbance, and eventually moves in the opposite direction without breakup \citep{Jullien2009}. We obtain the critical neck shape from the stop-flow simulations as shown in figure \ref{fig:5}(b) and determine the critical length by dividing the critical droplet shape from the top view in figure \ref{fig:5}(b) into five regions: the neck region (a), two plug-shaped regions (b), and two hemispherical head-ends (c). For different initial droplet lengths, neither the critical shape of the neck region nor the shape of the head-ends changes. The shortest breaking droplet should thus have a volume that comprises those of the neck region and two head-ends. A shorter droplet cannot reach the critical neck shape for breaking with a small $Ca$ number. According to the critical droplet shape in figure \ref{fig:5}(b), the critical length is approximately $3.07H$. This length is close to $\pi H$, which is used in \citep{Link2004} to predict the boundary between no breakup and breakup according to the Rayleigh-Plateau stability criterion. The viscosity ratio of dispersed and continuous phases alters the critical droplet length from approximately $2.62H$ to $4.20H$ as the viscosity ratio varies from 0.11 to 1.67 (Jullien, Ching et al. 2009), suggesting that the viscosity ratio may influence the critical shape of the neck region through the asymmetric flow field around the neck, including the leakage flow at the corners. Our results prove that the critical condition for breakup at low $Ca$ numbers is based on geometry and provide physical reasons for geometric conditions. The critical length for the axisymmetric configuration is determined from the theoretical prediction in figure \ref{fig:3} to be approximately $4.54R$ or $2.27D$ (with $D = 2R$). The critical length for the cross-flowing configuration with rectangular microchannels shown in figure \ref{fig:5}(a) is determined from numerical simulations to be approximately $2.28H$. These two critical lengths agree with each other. Naturally, the critical length can be used to predict the minimum droplet size that can be generated under low $Ca$ conditions.

\section{Concluding remarks}
The present study highlights the essential role of capillarity and ultimately illuminates the origins of capillary instability of confined fluid threads. Distinct from an unbounded fluid thread, a confined fluid thread is stabilized by its internal pressure and destabilized by the gradient of capillary pressure induced by confinements or local capillary pressure that exceeds this internal pressure. Revealed mechanisms not only explain the underlying physics but also can be used to invent new configurations of microfluidic devices that can further improve the efficiency of droplet/bubble generation.

\section*{Acknowledgements}
This work was partly supported by the National Natural Science Foundation of China (Grant Nos. 11772343, 11572334, 91543125, and 11402274), the CAS Key Research Program of Frontier Sciences (QYZDB-SSW-JSC036), the CAS Strategic Priority Research Program (XDB22040403), and the Beijing Institute of Technology Research Fund Program for Young Scholars.

\bibliographystyle{jfm}
\bibliography{jfm-instructions}

\begin{thebibliography}{34}
\expandafter\ifx\csname natexlab\endcsname\relax\def\natexlab#1{#1}\fi
\def\au#1{#1} \def\ed#1{#1} \def\yr#1{#1}\def\at#1{#1}\def\jt#1{\textit{#1}}
  \def\bt#1{#1}\def\bvol#1{\textbf{#1}} \def\vol#1{#1} \def\pg#1{#1}
  \def\publ#1{#1}\def\arxiv#1{#1}\def\org#1{#1}\def\st#1{\textit{#1}}

\bibitem[Anna(2016)]{Anna2016}
{\sc \au{Anna, S.~L.}} \yr{2016}  \at{Droplets and bubbles in microfluidic
  devices}.  \jt{Annu. Rev. Fluid Mech.}  \bvol{48},  \pg{285--309}.

\bibitem[Beresnev {\em et~al.\/}(2009)Beresnev, Li \& Vigil]{Beresnev2009}
{\sc \au{Beresnev, I.~A.}, \au{Li, W.~Q.} \& \au{Vigil, R.~D.}} \yr{2009}
  \at{Condition for break-up of non-wetting fluids in sinusoidally constricted
  capillary channels}.  \jt{Transport Porous Med.}  \bvol{80}~(3),
  \pg{581--604}.

\bibitem[Bhat {\em et~al.\/}(2010)Bhat, Appathurai, Harris, Pasquali, McKinley
  \& Basaran]{Bhat2010}
{\sc \au{Bhat, P.~P.}, \au{Appathurai, S.}, \au{Harris, M.~T.}, \au{Pasquali,
  M.}, \au{McKinley, G.~H.} \& \au{Basaran, O.~A.}} \yr{2010}  \at{Formation of
  beads-on-a-string structures during break-up of viscoelastic filaments}.
  \jt{Nat. Phys.}  \bvol{6}~(8),  \pg{625--631}.

\bibitem[Blanchette \& Bigioni(2006)]{Blanchette2006}
{\sc \au{Blanchette, F.} \& \au{Bigioni, T.~P.}} \yr{2006}  \at{Partial
  coalescence of drops at liquid interfaces}.  \jt{Nat. Phys.}  \bvol{2}~(4),
  \pg{254--257}.

\bibitem[Castrejon-Pita {\em et~al.\/}(2015)Castrejon-Pita, Castrejon-Pita,
  Thete, Sambath, Hutchings, Hinch, Lister \& Basaran]{CastrejonPita2015}
{\sc \au{Castrejon-Pita, J.~R.}, \au{Castrejon-Pita, A.~A.}, \au{Thete, S.~S.},
  \au{Sambath, K.}, \au{Hutchings, I.~M.}, \au{Hinch, J.}, \au{Lister, J.~R.}
  \& \au{Basaran, O.~A.}} \yr{2015}  \at{Plethora of transitions during breakup
  of liquid filaments}.  \jt{Proc. Natl. Acad. Sci. U.S.A.}  \bvol{112}~(15),
  \pg{4582--4587}.

\bibitem[Chen {\em et~al.\/}(2013)Chen, Ma, Yang \& Popinet]{Chen2013}
{\sc \au{Chen, X.}, \au{Ma, D.}, \au{Yang, V.} \& \au{Popinet, S.}} \yr{2013}
  \at{High-fidelity simulations of impinging jet atomization}.  \jt{Atomization
  Sprays}  \bvol{23}~(12),  \pg{1079--1101}.

\bibitem[Chen {\em et~al.\/}(2014)Chen, Xue, Zhang, Hu, Jiang \& Sun]{Chen2014}
{\sc \au{Chen, X.}, \au{Xue, C.}, \au{Zhang, L.}, \au{Hu, G.}, \au{Jiang, X.}
  \& \au{Sun, J.}} \yr{2014}  \at{Inertial migration of deformable droplets in
  a microchannel}.  \jt{Phys. Fluids}  \bvol{26}~(11),  \pg{25}.

\bibitem[Chen \& Yang(2014)]{ChenYang2014}
{\sc \au{Chen, X.} \& \au{Yang, V.}} \yr{2014}  \at{Thickness-based adaptive
  mesh refinement methods for multi-phase flow simulations with thin regions}.
  \jt{J. Comput. Phys.}  \bvol{269},  \pg{22--39}.

\bibitem[Dangla {\em et~al.\/}(2013)Dangla, Fradet, Lopez \&
  Baroud]{Dangla2013}
{\sc \au{Dangla, R.}, \au{Fradet, E.}, \au{Lopez, Y.} \& \au{Baroud, C.~N.}}
  \yr{2013}  \at{The physical mechanisms of step emulsification}.  \jt{J. Phys.
  D: Appl. Phys.}  \bvol{46}~(11),  \pg{114003}.

\bibitem[De~Menech {\em et~al.\/}(2008)De~Menech, Garstecki, Jousse \&
  Stone]{DeMenech2008}
{\sc \au{De~Menech, M.}, \au{Garstecki, P.}, \au{Jousse, F.} \& \au{Stone,
  H.~A.}} \yr{2008}  \at{Transition from squeezing to dripping in a
  microfluidic t-shaped junction}.  \jt{J. Fluid Mech.}  \bvol{595},
  \pg{141--161}.

\bibitem[Delaunay(1841)]{Delaunay1841}
{\sc \au{Delaunay, C.~H.}} \yr{1841}  \at{Sur la surface de revolution dont la
  courbure moyenne est constante}.  \jt{J. Math. Pures Appl.}  \pg{pp.
  309--314}.

\bibitem[Dollet {\em et~al.\/}(2008)Dollet, van Hoeve, Raven, Marmottant \&
  Versluis]{Dollet2008}
{\sc \au{Dollet, B.}, \au{van Hoeve, W.}, \au{Raven, J.~P.}, \au{Marmottant,
  P.} \& \au{Versluis, M.}} \yr{2008}  \at{Role of the channel geometry on the
  bubble pinch-off in flow-focusing devices}.  \jt{Phys. Rev. Lett.}
  \bvol{100}~(3),  \pg{034504}.

\bibitem[Duclaux {\em et~al.\/}(2006)Duclaux, Clanet \& Quere]{Duclaux2006}
{\sc \au{Duclaux, V.}, \au{Clanet, C.} \& \au{Quere, D.}} \yr{2006}  \at{The
  effects of gravity on the capillary instability in tubes}.  \jt{J. Fluid
  Mech.}  \bvol{556},  \pg{217--226}.

\bibitem[Eggers {\em et~al.\/}(2007)Eggers, Fontelos, Leppinen \&
  Snoeijer]{Eggers2007}
{\sc \au{Eggers, J.}, \au{Fontelos, M.~A.}, \au{Leppinen, D.} \& \au{Snoeijer,
  J.~H.}} \yr{2007}  \at{Theory of the collapsing axisymmetric cavity}.
  \jt{Phys. Rev. Lett.}  \bvol{98}~(9),  \pg{094502}.

\bibitem[Eggers \& Villermaux(2008)]{Eggers2008}
{\sc \au{Eggers, J.} \& \au{Villermaux, E.}} \yr{2008}  \at{Physics of liquid
  jets}.  \jt{Rep. Prog. Phys.}  \bvol{71}~(3),  \pg{036601}.

\bibitem[Garstecki {\em et~al.\/}(2005)Garstecki, Stone \&
  Whitesides]{Garstecki2005}
{\sc \au{Garstecki, P.}, \au{Stone, H.~A.} \& \au{Whitesides, G.~M.}} \yr{2005}
   \at{Mechanism for flow-rate controlled breakup in confined geometries: a
  route to monodisperse emulsions}.  \jt{Phys. Rev. Lett.}  \bvol{94}~(16),
  \pg{164501}.

\bibitem[Ghorbanian {\em et~al.\/}(2010)Ghorbanian, Qasaimeh \&
  Juncker]{Ghorbanian2010}
{\sc \au{Ghorbanian, S.}, \au{Qasaimeh, M.~A.} \& \au{Juncker, D.}} \yr{2010}
  Rapid prototyping of branched microfluidics in pdms using capillaries.

\bibitem[Gordillo {\em et~al.\/}(2005)Gordillo, Sevilla, Rodriguez-Rodriguez \&
  Martinez-Bazan]{Gordillo2005}
{\sc \au{Gordillo, J.~M.}, \au{Sevilla, A.}, \au{Rodriguez-Rodriguez, J.} \&
  \au{Martinez-Bazan, C.}} \yr{2005}  \at{Axisymmetric bubble pinch-off at high
  reynolds numbers}.  \jt{Phys. Rev. Lett.}  \bvol{95}~(19),  \pg{194501}.

\bibitem[Hoang {\em et~al.\/}(2013)Hoang, Portela, Kleijn, Kreutzer \& van
  Steijn]{Hoang2013}
{\sc \au{Hoang, D.~A.}, \au{Portela, L.~M.}, \au{Kleijn, C.~R.}, \au{Kreutzer,
  M.~T.} \& \au{van Steijn, V.}} \yr{2013}  \at{Dynamics of droplet breakup in
  a t-junction}.  \jt{J. Fluid Mech.}  \bvol{717},  \pg{R4}.

\bibitem[van Hoeve {\em et~al.\/}(2011)van Hoeve, Dollet, Versluis \&
  Lohse]{VanHoeve2011}
{\sc \au{van Hoeve, W.}, \au{Dollet, B.}, \au{Versluis, M.} \& \au{Lohse, D.}}
  \yr{2011}  \at{Microbubble formation and pinch-off scaling exponent in
  flow-focusing devices}.  \jt{Phys. Fluids}  \bvol{23}~(9),  \pg{092001}.

\bibitem[van Hoeve {\em et~al.\/}(2010)van Hoeve, Gekle, Snoeijer, Versluis,
  Brenner \& Lohse]{VanHoeve2010}
{\sc \au{van Hoeve, W.}, \au{Gekle, S.}, \au{Snoeijer, J.~H.}, \au{Versluis,
  M.l}, \au{Brenner, M.~P.} \& \au{Lohse, D.}} \yr{2010}  \at{Breakup of
  diminutive rayleigh jets}.  \jt{Phys. Fluids}  \bvol{22}~(12),  \pg{122003}.

\bibitem[Jullien {\em et~al.\/}(2009)Jullien, Ching, Cohen, Menetrier \&
  Tabeling]{Jullien2009}
{\sc \au{Jullien, M.~C.}, \au{Ching, M. J. T.~M.}, \au{Cohen, C.},
  \au{Menetrier, L.} \& \au{Tabeling, P.}} \yr{2009}  \at{Droplet breakup in
  microfluidic t-junctions at small capillary numbers}.  \jt{Phys. Fluids}
  \bvol{21}~(7),  \pg{072001}.

\bibitem[Link {\em et~al.\/}(2004)Link, Anna, Weitz \& Stone]{Link2004}
{\sc \au{Link, D.~R.}, \au{Anna, S.~L.}, \au{Weitz, D.~A.} \& \au{Stone,
  H.~A.}} \yr{2004}  \at{Geometrically mediated breakup of drops in
  microfluidic devices}.  \jt{Phys. Rev. Lett.}  \bvol{92}~(5),  \pg{054503}.

\bibitem[Olbricht(1996)]{Olbricht1996}
{\sc \au{Olbricht, W.~L.}} \yr{1996}  \at{Pore-scale prototypes of multiphase
  flow in porous media}.  \jt{Annu. Rev. Fluid Mech}  \bvol{28}~(1),
  \pg{187--213}.

\bibitem[Plateau(1873)]{Plateau1873}
{\sc \au{Plateau, J.}} \yr{1873}  \at{Experimental and theoretical steady state
  of liquids subjected to nothing but molecular forces}.
  \jt{Gauthiers-Villars, Paris} .

\bibitem[Popinet(2009)]{Popinet2009}
{\sc \au{Popinet, S.}} \yr{2009}  \at{An accurate adaptive solver for
  surface-tension-driven interfacial flows}.  \jt{J. Comput. Phys.}
  \bvol{228}~(16),  \pg{5838--5866}.

\bibitem[Rayleigh(1878)]{Rayleigh1878}
{\sc \au{Rayleigh, Lord}} \yr{1878}  \at{On the instability of jets}.  \jt{P.
  Lond. Math. Soc.}  \bvol{1}~(1),  \pg{4--13}.

\bibitem[Roof(1970)]{Roof1970}
{\sc \au{Roof, J.~G.}} \yr{1970}  \at{Snap-off of oil droplets in water-wet
  pores}.  \jt{Soc. Petrol. Eng. J.}  \bvol{10}~(1),  \pg{85--90}.

\bibitem[Shui {\em et~al.\/}(2008)Shui, Mugele, van~den Berg \&
  Eijkel]{Shui2008}
{\sc \au{Shui, L.~L.}, \au{Mugele, F.}, \au{van~den Berg, A.} \& \au{Eijkel, J.
  C.~T.}} \yr{2008}  \at{Geometry-controlled droplet generation in head-on
  microfluidic devices}.  \jt{Appl. Phys. Lett.}  \bvol{93}~(15),  \pg{153113}.

\bibitem[Squires \& Quake(2005)]{Squires2005}
{\sc \au{Squires, T.~M.} \& \au{Quake, S.~R.}} \yr{2005}  \at{Microfluidics:
  Fluid physics at the nanoliter scale}.  \jt{Rev. Mod. Phys.}  \bvol{77}~(3),
  \pg{977--1026}.

\bibitem[van Steijn {\em et~al.\/}(2009)van Steijn, Kleijn \&
  Kreutzer]{VanSteijn2009}
{\sc \au{van Steijn, V.}, \au{Kleijn, C.~R.} \& \au{Kreutzer, M.~T.}} \yr{2009}
   \at{Flows around confined bubbles and their importance in triggering
  pinch-off}.  \jt{Phys. Rev. Lett.}  \bvol{103}~(21),  \pg{214501}.

\bibitem[Stone {\em et~al.\/}(1986)Stone, Bentley \& Leal]{Stone1986}
{\sc \au{Stone, H.~A.}, \au{Bentley, B.~J.} \& \au{Leal, L.~G.}} \yr{1986}
  \at{An experimental-study of transient effects in the breakup of viscous
  drops}.  \jt{J. Fluid Mech.}  \bvol{173},  \pg{131--158}.

\bibitem[Tagawa {\em et~al.\/}(2012)Tagawa, Oudalov, Visser, Peters, van~der
  Meer, Sun, Prosperetti \& Lohse]{Tagawa2012}
{\sc \au{Tagawa, Y.}, \au{Oudalov, N.}, \au{Visser, C.~W.}, \au{Peters, I.~R.},
  \au{van~der Meer, D.}, \au{Sun, C.}, \au{Prosperetti, A.} \& \au{Lohse, D.}}
  \yr{2012}  \at{Highly focused supersonic microjets}.  \jt{Phys. Rev. X}
  \bvol{2}~(3),  \pg{031002}.

\bibitem[Wong {\em et~al.\/}(1995)Wong, Radke \& Morris]{Wong1995}
{\sc \au{Wong, H.}, \au{Radke, C.~J.} \& \au{Morris, S.}} \yr{1995}  \at{The
  motion of long bubbles in polygonal capillaries. part 1. thin films}.  \jt{J.
  Fluid Mech.}  \bvol{292},  \pg{71--94}.

\end{thebibliography}

\end{document}